\documentclass[epj]{svjour}
\usepackage{latexsym}
\usepackage{graphics}
\usepackage{amsmath}
\begin{document}
\title{A quantum secret sharing scheme with verifiable function}
\author{Li-Juan Liu\inst{1} \and Zhi-Hui Li\inst{1}% etc
\thanks{\emph{e-mail:lizhihui@suun.edu.cn} }
\and Zhao-Wei Han\inst{1}
\and Dan-Li Zhi\inst{1}%
}                % Do not remove
\institute{$^{1}$ College of Mathematics and Information Science, Shaanxi Normal University, ${\rm Xi^{'}}$an 710119, P.R. China  }
\date{Received: date / Revised version: date}
\abstract{
In the $\left( {t,n} \right)$ threshold quantum secret
sharing scheme, it is difficult to ensure that internal participants are
honest. In this
paper, a verifiable $\left( {t,n} \right)$ threshold quantum secret sharing
scheme is designed combined with classical secret sharing scheme. First of
all, the distributor uses the asymmetric binary polynomials to generate the
shares and sends them to each participant. Secondly, the distributor sends
the initial quantum state with the secret to the first participant,
and each participant performs unitary operation that using the mutually
unbiased bases on the obtained $d$ dimension single bit quantum state ($d$
is a large odd prime number). In this process, distributor can randomly
check the participants, and find out the internal fraudsters by unitary
inverse operation gradually upward. Then the secret is reconstructed
after all other participants simultaneously public transmission. Security
analysis show that this scheme can resist both external and internal
attacks.}
\PACS{
       {00}{00} \and
       {20}{00} \and
       {42}{10}
       }
     % end of PACS codes
%} %end of abstract
%
\maketitle
\section{Introduction} \ \ \ \
In 1979, the secret sharing scheme was first proposed by Shamir \cite{RefA} and Blakely \cite{RefB}, which is an important technology to ensure the security and availability of confidential information. In addition, they are widely used as the components of various cryptographic protocols, such as threshold cryptography, attribute-based encryption and multi-party computing. In the $\left( {t,n} \right)$ threshold secret sharing scheme, the secret is divided into $n$ shares so that it can only be recovered with $t$ or more than $t$ shares, but fewer than $t$ shares cannot reveal any information of the secret. At present, the research of classical secret sharing scheme has become mature \cite{RefC,RefD,RefE}. However, most of the schemes have the following potential security hazard: it is impossible to check the honesty of internal participants in the secret recovery phase.
Therefore, the verifiable secret sharing (VSS) scheme was proposed by Chor
et al. \cite{RefF} in 1985. The purpose of the VSS scheme is to prevent participants
from providing wrong shares in the secret recovery phase. So far, more and
more theories of VSS \cite{RefG,RefH} have been put forward. However, all of the VSS
schemes are based on the assumption of computational complexity, namely security
is conditional. With the improvement of computing capabilities and
algorithms, especially the emergence of quantum algorithms \cite{RefI}, the security
of classical cryptography is facing severe challenges. In addition, as the
extension of classical secret sharing scheme in the quantum field, the
research of quantum secret sharing scheme plays an important role.

In 1999, the quantum secret sharing (QSS) scheme was first proposed by
Hillery, Buzek and Berthiaume \cite{RefJ},which attracted widely attention. QSS
scheme still adopts the secret system of classical secret sharing scheme,
which uses the quantum state as the encoding carrier of secret information.
QSS scheme can be divided into two categories starting from the type of
shared information: 1) QSS scheme of sharing classical information \cite{RefK,RefL,RefM};
2) QSS scheme of sharing quantum information(quantum state) \cite{RefN,RefO,RefP,RefQ}. The
former presents various characteristics in the scheme design, while the
latter is mostly realized by means of quantum entanglement swapping and
quantum teleportation.
Many QSS schemes are $(n,n)$ threshold \cite{RefR,RefS,RefT} that need all
participants to reconstruct the secret together. In order to improve the
flexibility and practicability, the $(t,n)$ threshold quantum
secret sharing (TQSS) schemes \cite{RefN,RefU,RefV,RefW} were proposed. These TQSS schemes
only aim at how to make $t$ shares or more than $t$ shares reconstruct secret through security channel. However,
there may be dishonest participants to provide wrong shares which lead to
the errors of recovered secret in real life. Thus the verifiable $(t,n)$ threshold quantum secret sharing (VTQSS) scheme was proposed to
verify the shares. In 2011, the VTQSS scheme was proposed by Yang et
al. \cite{RefX}. The security of it was analyzed by Song and Liu \cite{RefY}, and they found
it could not prevent forgery attack of participants. In 2016, Qin and
Dai \cite{RefZ} proposed a VTQSS scheme using $d$ dimension Bell
state. With the higher and higher requirements for verifiability, the research on VTQSS scheme \cite{Refa} is gradually in-depth.

Lu et al. \cite{Refb} proposed the VTQSS scheme based on the threshold secret
sharing scheme of shamir which using the $d$ dimension
single bit quantum state. In this scheme, a detection particle is added to
detect external attack and internal fraudsters by using the equality
relationship between two secrets. However, the number used to reconstruct
the secrets is randomly selected by the participants, the distributor cannot
identify them. If there are dishonest internal participants who provide
wrong shares in the secret sharing phase, the scheme is destroyed but the
internal attacks cannot be detected.

In this paper, we propose a VTQSS scheme based on $d$
dimension single bit quantum state, and $d$ is a large odd
prime number. In this scheme, the distributor generates shares and
distributes them to each participant in the classical secret distribution
phase. In the secret sharing phase, after all participants perform unitary
operations continuously on the quantum state prepared by the distributor,
the last participant uses the measurement basis sent by the distributor to
obtain the result, which is publicly transmitted to the distributor and all
other participants simultaneously. Then the other participants transmit the
information they have to the distributor and the participants who need to
recover the secret simultaneously, so the secret is reconstructed. The
advantages of the scheme are as follows:

\begin{enumerate}
\item[1)]The subshares used to reconstruct the secrets is generated by asymmetric binary polynomials, which are owned by both the distributor and the participant, and can be detected at any time.
\item[2)]The distributor can find out the scheme errors in time by checking randomly, and can find out the fraudsters and eliminate them by upward step-by-step inspection, which use the unitary inverse operation. Thus the constant waste of resources can be avoided.
\item[3)] Public transmission simultaneously can not only prevent participants from sending wrong results, but also prevent other participants from pretending the participant to fraud.
\item[4)] The TQSS scheme is more flexible and applicable than the $(n,n)$ threshold QSS scheme.
\end{enumerate}

The structure of this paper is as follows. In the section two, the classical secret sharing scheme which is based on the asymmetric binary polynomials, mutually unbiased bases and simultaneous public transmission is reviewed. In the section three, the improved VTQSS scheme is introduced. In the section four, the security of this scheme is analyzed. In the section five, namely the last part, this scheme is summarized.

\section{Basic knowledge} \ \ \ \
In this section, we will introduce the basic knowledge used in the scheme
design, including the concept of protected secret sharing scheme, mutually
unbiased bases and simultaneous public transmission.

\subsection{Classical secret sharing scheme based on asymmetric binary polynomials} \ \ \ \
\label{sec:1}
Lein et al. proposed the protected secret sharing (PSS) scheme in 2017. The
scheme is designed by using asymmetric binary polynomials $F(x,y)$, where $F(x,y)$ is with degree at most $t-1$ in $x$ and with degree at most $h-1$ in $y$. It can be expressed as
\begin{equation}
\label{eq1}
F(x,y)=a_{0,0}+a_{1,0}x+a_{0,1}y+\cdots+a_{t-1,h-1}x^{t-1}y^{h-1}.
\end{equation}
where $a_{i,j}  \in D$, $\forall i,j \in \left[ {0,t - 1} \right]$
and the coefficient satisfies $a_{i,j}  \ne a_{j,i}$, $
\forall i,j \in \left[ {0,t - 1} \right]$. This kind of polynomials is called
as asymmetrical binary polynomial.

In the PSS scheme, the distributor uses an asymmetric binary polynomial
$F(x,y)$ to generate a pair of shares $s_{i}^{(1)}(y)=F(x_{i},y)$ and
$s_{i}^{(2)}(x)=F(x,x_{i})$ for shareholders, where $i=1,2,\cdots,n$. The shares
$F(x_{i},y)$ and $F(x,x_{i})$ are all univariate polynomials,
where $F(x_{i},y)$ is with degree at most $h-1$ and
$F(x,x_{i})$ is with degree at most $t-1$, and $F(x_{i},x_{j})\neq F(x_{j},x_{i})$. A pairwise keys can be established between the shareholders
$U_{i}$ and $U_{j}$:
\[
k_{i,j}=s_{i}^{(1)}(x_{j})=s_{j}^{(2)}(x_{i})=F(x_{i},x_{j}),
\]
\[
k_{j,i}=s_{i}^{(2)}(x_{j})=s_{j}^{(1)}(x_{i})=F(x_{j},x_{i}).
\]

Therefore, a pairwise shared key can be established between two shareholders
to ensure that the reconstructed secret is not obtained by the
nonshareholders by using asymmetric binary polynomial.

\subsection{Mutually unbiased bases} \ \ \ \
In many quantum information processing, mutually unbiased bases (MUBs) plays
an important role. The knowledgeable of MUBs is given as follows.

\noindent \textbf{Definition 1} Assuming that

\noindent $B_0  = \left\{ {\left| {\varphi _0^{\left( 1 \right)} } \right\rangle ,\left| {\varphi _0^{\left( 2 \right)} } \right\rangle , \cdots ,\left| {\varphi _0^{\left( d \right)} } \right\rangle } \right\}$
and

\noindent $B_1  = \left\{ {\left| {\varphi _1^{\left( 1 \right)} } \right\rangle ,\left| {\varphi _1^{\left( 2 \right)} } \right\rangle , \cdots ,\left| {\varphi _1^{\left( d \right)} } \right\rangle } \right\}$ are two sets of orthonormal bases in $d$ dimension space, if
they satisfy
\[
\left| {\left\langle {{\varphi _0^{\left( l \right)} }}
 \mathrel{\left | {\vphantom {{\varphi _0^{\left( l \right)} } {\varphi _1^{\left( j \right)} }}}
 \right. \kern-\nulldelimiterspace}
 {{\varphi _1^{\left( j \right)} }} \right\rangle } \right| = \frac{1}{{\sqrt d }}.
\]
they are said to be unbiased.

If any two sets of orthonormal bases $\left\{ {B_0 ,B_1 , \cdots ,B_m } \right\}$
in $C^{d}$ space are unbiased, then the set is called unbiased base set.

It is known from the literature \cite{Refc,Refd} that when the dimension of quantum
system $d$ is an odd prime number, at least $d+1$ MUBs can be found. In particular, the computation basis is
expressed as $\{|k\rangle|k\in D\}$, where $D=\{0,1,\cdots,d-1\}$. For the sake of
consistency, this scheme is limited $d$ to an odd prime
number. In addition to the computation basis, the remaining $d$ MUBs can be expressed as:
\[
\left| {\phi _l^{\left( j \right)} } \right\rangle  = \frac{1}{{\sqrt d }}\sum\limits_{k = 0}^{d - 1} {\omega ^{k\left( {l + jk} \right)} \left| k \right\rangle }
\]

\noindent where $\omega  = e^{{{2\pi i} \mathord{\left/
 {\vphantom {{2\pi i} d}} \right.
 \kern-\nulldelimiterspace} d}}$, $j \in D$ represent the number of unbiased bases, $l \in D$ enumerate the number of vectors for the given base., These unbiased bases satisfy:
 \[
|{\langle\varphi_{l}^{\left(j\right)}|\varphi_{l^{'}}^{\left(j^{'}\right)}\rangle}|= \frac{1}{{\sqrt d }}
 \]
where $j\neq j^{'}$

We can see from the formula $\left| {\varphi _l^{\left( j \right)} } \right\rangle  = \frac{1}{{\sqrt d }}\sum\limits_{k = 0}^{d - 1} {\omega ^{k\left( {l + jk} \right)} \left| k \right\rangle }$ that $ B = \left\{ {B_0 ,B_1 , \cdots B_m , \cdots B_d } \right\}$ is a set of MUBs:

when $j=0$,

$B_0  = \left\{ {\left| {\varphi _0^{\left( 0 \right)} } \right\rangle ,\left| {\varphi _1^{\left( 0 \right)} } \right\rangle , \cdots ,\left| {\varphi _l^{\left( 0 \right)} } \right\rangle , \cdots \left| {l \in D} \right.} \right\}$;

when $j=1$,

$B_1  = \left\{ {\left| {\varphi _0^{\left( 1 \right)} } \right\rangle ,\left| {\varphi _1^{\left( 1 \right)} } \right\rangle , \cdots ,\left| {\varphi _l^{\left( 1 \right)} } \right\rangle , \cdots \left| {l \in D} \right.} \right\}$;
\[
\cdots\cdots
\]

when $j=m$,

$ B_m  = \left\{ {\left| {\varphi _0^{\left( m \right)} } \right\rangle ,\left| {\varphi _1^{\left( m \right)} } \right\rangle , \cdots ,\left| {\varphi _l^{\left( m \right)} } \right\rangle , \cdots \left| {l \in D} \right.} \right\}$;
\[
\cdots\cdots
\]

when $j=d$, we can make

$B_d  = \left\{ {\left| 0 \right\rangle ,\left| 1 \right\rangle , \cdots ,\left| l \right\rangle , \cdots \left| {l \in D} \right.} \right\}$.

The coding operations in the literature \cite{RefR} are composed of two unitary
operators $X_{d}$ and $Y_{d}$, where:
\[
X_d  = \sum\limits_{n = 0}^{d - 1} {\omega ^n \left| n \right\rangle \left\langle n \right|},
\]
\[
Y_d  = \sum\limits_{n = 0}^{d - 1} {\omega ^{n^2 } \left| n \right\rangle \left\langle n \right|}.
\]

\noindent \textbf{Lemma 1} $D$ is a finite field. From the formula $
\left| {\varphi _l^{\left( j \right)} } \right\rangle  = \frac{1}{{\sqrt d }}\sum\limits_{k = 0}^{d - 1} {\omega ^{k\left( {l + jk} \right)} \left| k \right\rangle }$, we can see that the MUBs has the following properties:

\begin{enumerate}
\item[1)] when the unitary operator $X_{d}$ is applied to particle $\left| {\varphi _l^{\left( j \right)} } \right\rangle$, its subscript will change, namely $X_d^x \left| {\varphi _l^{\left( j \right)} } \right\rangle  = \left| {\varphi _{l + x}^{\left( j \right)} } \right\rangle$.
\item[2)] when the unitary operator $Y_{d}$
 is applied to particle $\left| {\varphi _l^{\left( j \right)} } \right\rangle$, its superscript will change, namely $
Y_d^y \left| {\varphi _l^{\left( j \right)} } \right\rangle  = \left| {\varphi _l^{\left( {j + y} \right)} } \right\rangle$.
\end{enumerate}

\noindent \textbf{Proof} 1)
\begin{equation}
\begin{aligned}
X_d^x \left| {\varphi _l^{\left( j \right)} } \right\rangle&= \left( {\sum\limits_{n = 0}^{d - 1} {\omega ^{xn} \left| n \right\rangle \left\langle n \right|} } \right)\left( {\frac{1}{{\sqrt d }}\sum\limits_{k = 0}^{d - 1} {\omega ^{k\left( {l + jk} \right)} \left| k \right\rangle } } \right)\nonumber\\
&= \left( {\frac{1}{{\sqrt d }}\sum\limits_{k = 0}^{d - 1} {\omega ^{k\left[ {\left( {l + x} \right) + jk} \right]} \left| k \right\rangle } } \right)\nonumber\\
&= \left| {\varphi _{\left( {l + x} \right)}^{\left( j \right)} } \right\rangle.\
\end{aligned}
\end{equation}

The following 2) can be proved by the same way.

For $\forall x,y \in D$, a unitary matrix can be constructed so that $\left| {\varphi _l^{\left( j \right)} } \right\rangle$
can be transformed into $\left| {\varphi _{l + x}^{\left( {j + y} \right)} } \right\rangle$. That is to map the
elements of $B$ into $B$. Let the unitary
matrix $U_{x,y}  = X_x^d Y_y^d$, then we have $U_{x,y} \left| {\varphi _l^{\left( j \right)} } \right\rangle  = \left| {\varphi _{l + x}^{\left( {j + y} \right)} } \right\rangle$.

\subsection{Simultaneous public transmission}
\ \ \ \ The so-called simultaneous public transmission means that a participant
sends a message to different people through secret channels at the same
time, which is the same as the concept of mail CC. It can realize:

\begin{enumerate}
\item[1)]The sender sends the information simultaneously and the receiver receives it at the same time. It can prevent some receivers from receiving the correct information in advance, but forge the sender to send the wrong information to other receivers.
\item[2)]That is to send the same message. It can prevent the sender from sending different information to different receivers, which will result in wrong information.
\end{enumerate}

\section{Scheme description} \ \ \ \
In this section, we present a $\left( {t,n} \right)$ threshold secret sharing
scheme using the single bit quantum state, which includes two parts: the
classical secret distribution phase and the secret sharing phase. Alice is
the distributor and  ${\rm Bob_i}$$\left( {i = 1,2, \cdots ,m} \right)$
are the participants who has the
only public identity $x_i \left( {i \in D} \right)$, that is $x_i  \ne x_j \left( {i \ne j} \right)$.

\subsection{Classical secret distribution phase} \ \ \ \
At first, Alice chooses a random asymmetric binary polynomial:
\[
F\left( {x,y} \right) = a_{0,0}  + a_{1,0} x + a_{0,1} y +  \cdots  + a_{t - 1,h - 1} x^{t - 1} y^{h - 1}.
\]

\noindent which need to meet $h > t\left( {t - 1} \right)$.

\begin{enumerate}
\item[1)] Alice calculates a pair of shares $s_i^{\left( 1 \right)} \left( y \right) = F\left( {x_i ,y} \right)$
 and $s_i^{\left( 2 \right)} \left( x \right) = F\left( {x,x_i } \right)$, and sends $\left\{ {s_i^{\left( 1 \right)} \left( y \right),s_i^{\left( 2 \right)} \left( x \right)} \right\}$ through the secret channel to the ${\rm Bob_i}$.
\item[2)] Participants ${\rm Bob_i}$ and ${\rm Bob_j}$(assume $i<j$) can compute a pairwise shared key:
\end{enumerate}
\[
\left\{ \begin{array}{l}
 k_{i,j}  = s_i^{\left( 1 \right)} \left( {x_j } \right) = s_j^{\left( 2 \right)} \left( {x_i } \right) = F\left( {x_i ,x_j } \right) \\
 k_{j,i}  = s_i^{\left( 2 \right)} \left( {x_j } \right) = s_j^{\left( 1 \right)} \left( {x_i } \right) = F\left( {x_j ,x_i } \right) \\
 \end{array}. \right.
\]

\subsection{Secret sharing phase} \ \ \ \
At first, the distributor Alice prepares the quantum state $
\left| \Phi  \right\rangle  = \left| {\varphi _0^{\left( 0 \right)} } \right\rangle  = \frac{1}{{\sqrt d }}\sum\limits_{j = 1}^d {\left| j \right\rangle }
$ and the secret $S \in D$.

\begin{enumerate}
\item[1)] Alice performs unitary operation $U_{p_0 ,q_0 }  = X_d^{p_0 } Y_d^{q_0 }$ on $\left| \Phi  \right\rangle$, that is $
U_{p_0 ,q_0 } \left| \Phi  \right\rangle  = \left| \Phi  \right\rangle _0  = \left| {\varphi _{p_0 }^{q_0 } } \right\rangle$, where $
p_0  = S$, and $q_0$ is arbitrary value.
\item[2)] Assuming that Alice needs to share the secret among $m$ participants $\left\{ {{\rm Bob_i} ,i = 1,2, \cdots ,m} \right\}$, she will send $
\left| \Phi  \right\rangle _0$ to ${\rm Bob_1}$ at first.
\item[i)] After ${\rm Bob_1}$ receiving $\left| \Phi  \right\rangle _0$, he will perform unitary operation $U_{p_1 ,q_1 }$ on it, where $
p_1  = k_{1,2} ,q_1  = k_{2,1}$. Then the quantum state $\left| \Phi  \right\rangle _0$
 changes to $\left| \Phi  \right\rangle _1  = \left| {\varphi _{p_0  + p_1 }^{q_0  + q_1 } } \right\rangle$, ${\rm Bob_1}$ sends $\left| \Phi  \right\rangle _1$
 to ${\rm Bob_2}$.
\item[ii)] Participant ${\rm Bob_j} ,j = 1,2, \cdots m$
 repeats the operation that $ {\rm Bob_1 }$
 conducts in i). That is, $ {\rm Bob_j}$
 performs unitary operation $U_{p_j ,q_j }$
 on $\left| \Phi  \right\rangle _{j - 1}$, then he gets the quantum state $
\left| \Phi  \right\rangle _j  = \left| {\varphi _{\sum\limits_{r = 0}^j {p_r } }^{\sum\limits_{r = 0}^j {q_r } } } \right\rangle$, where $
p_j  = k_{j,j + 1} $, $q_j  = k_{j + 1,j}$, $p_{j},q_j  \in D$. ${\rm Bob_j}$
 will send $\left| \Phi  \right\rangle _j$
 to the next participant ${\rm Bob_{j+1}}$,
$j = 2,3, \cdots m - 1$.
\item[3)] In the process of 2), Alice randomly checks $
{\rm Bob_u} $

\noindent $\left( {0 \le u \le m} \right)$. She sends $
\sum\limits_{i = 0}^u {q_i }$
 to ${\rm Bob_u}$ by using $c_u  = E_{s_u^{\left( 1 \right)} \left( 0 \right)} \left( {\sum\limits_{i = 0}^u {q_i } } \right)$ encryption. After
 ${\rm Bob_u}$ receiving $c_u$, $\sum\limits_{i = 0}^u {q_i }  = j$ is obtained by using
 $D_{s_u^{\left( 1 \right)} \left( 0 \right)} \left( {c_i } \right)$ decryption.
 He measures $\left| \Phi  \right\rangle _u$ to get $l_u$ by using the measurement basis
 $\left\{ {\left| {\varphi _l^{\left( j \right)} } \right\rangle } \right\}_l$, then he encrypts and sends it to Alice.
 If Alice check $l_u  = \sum\limits_{i = 0}^u {p_i }$ is satisfied, there are no internal fraudsters in the previous $u$ participants. Then the scheme can continue.
\item[i)] If Alice find out the internal fraudster is existence, the participant ${\rm Bob_u}$ performs the unitary inverse operation $U_{ - p_u , - q_u }
$ on $\left| \Phi  \right\rangle _u$, then he sends the result
$|\Phi\rangle_{u-1}^{'}$ to ${\rm Bob_{u - 1}}$. The above operation is repeated again, Alice check whether
$l_{u - 1}  = \sum\limits_{i = 0}^{u - 1} {p_i }$ is satisfied. If it is satisfied, ${\rm Bob_u}$ is the internal fraudsters and he will be eliminated.
\item[ii)] If it is not satisfied, the above i) operation is repeated again, and check it upward in turn until the internal fraudsters is found out and eliminated.
\item[4)] The last participant ${\rm Bob_m}$ gets the quantum state $\left| \Phi  \right\rangle _m$ by unitary operation, and he chooses $
\left\{ {\left| {\varphi _l^{\left( j \right)} } \right\rangle } \right\}_l$ as the measurement basis to measure the quantum state and gets the result $R$. After ${\rm Bob_m}$ uses $c_{k_{i,m} }  = E_{k_{i,m} } \left( R \right)$ to encrypt $R$, he simultaneous public transmits it to Alice and ${\rm Bob_i}$ through the security channel. Then they can get $R$ after using $D_{k_{i,m} } \left( {c_{k_{_{i,m} } } } \right)$ to decrypt. In this part, the measurement basis $
j = \sum\limits_{i = 1}^m {q_i }$ is sent to ${\rm Bob_m}$ through the the same way as 3) by Alice.
\item[5)] After ${\rm Bob_j} \left( {j \ne i} \right)$ uses $c_{k_{j,i} }  = E_{k_{j,i} } \left( {p_j } \right)
$ to encrypt $p_j$, he simultaneous public transmits it to Alice and ${\rm Bob_i}$ through the security channel. Then they can get $P_j$ after using
$D_{k_{j,i} } \left( {c_{k_{j,i} } } \right)$ to decrypt. When Alice and ${\rm Bob_i}$
$\left( {i = 1,2, \cdots ,m;i \ne j} \right)$ get $R$ and all $
P_j \left( {j \ne i} \right)$, Alice checks whether the number obtained is correct.
\item[6)] If Alice checks it is correct, ${\rm Bob_i} \left( {i = 1,2, \cdots ,m} \right)$ can reconstruct the secret
\[
P_0  = R - \sum\limits_{j = 1}^m {p_j }.
\]
\end{enumerate}

Otherwise, Alice will terminate the scheme, remove the participants who sent the wrong number more than twice, and start it again.

\section{Security analysis} \ \ \ \
In this section, the completeness and safety of the scheme will be analyzed.
The security analysis includes external attack and internal attack.
\subsection{Completeness analysis} \ \ \ \
We will prove the completeness of the scheme next. In the $\left( {t,n} \right)$ threshold secret sharing scheme, the secret is divided into
$n$ shares. Only through $t$ or more than
$t$ shares can the secret be reconstructed, but less than
$t$ shares cannot recover any information of the secret. It
means the scheme satisfies the completeness.

\begin{enumerate}
\item[1)] In the classical secret distribution phase, if the asymmetric binary polynomial $F\left( {x,y} \right)$ satisfies
$h > t\left( {t - 1} \right)$, then $t$ shares or more than $t$ shares can reconstruct the secret, but less than $t$ shares cannot get any information.
\end{enumerate}
\textbf{Proof} \ \ For $h > t\left( {t - 1} \right)$, because $F\left( {x,y} \right)$ is an asymmetric
binary polynomial, where the degree of $x$ is $t-1$ and the degree of $y$ is $h-1$,
it contains $th$ different coefficients. In this proposed
scheme, each share $\left\{ {\left. {s_i^1 \left( y \right),s_i^2 \left( x \right)} \right\}} \right.$ contains two univariate polynomials,
where the degree of $y$ is $h-1$ and the
degree of $x$ is $t-1$. In other words, each
shareholder can use its shares to establish at most $t+h$
linearly independent equations according to the coefficients of the binary
polynomial $F\left( {x,y} \right)$. When there are $t-1$
shareholders merging with their shares, they can establish a total of $\left( {t + h} \right)\left( {t - 1} \right)$ linearly independent equations. If the number of the coefficients of the binary polynomial $F\left( {x,y} \right)$ is larger than the
number of equations of the combined shareholders, that is, $th > \left( {t + h} \right)\left( {t - 1} \right)$. $t-1$ dishonest shareholders cannot recover
$F\left( {x,y} \right)$. As a result, they cannot get any secret information.
Thus $h > t\left( {t - 1} \right)$ can ensure that less than $t$
 shares cannot disclose any secret information.

\begin{enumerate}
\item[2)] In the secret sharing phase, it can be seen from $S = p_0  = R - \sum\limits_{j = 1}^m {p_j }$ that we need $m$ participants cooperate to reconstruct secret.
\end{enumerate}

\subsection{Security analysis}
\begin{enumerate}
\item[1)] External attack
\end{enumerate}
\begin{enumerate}
\item[i)] Intercept-and-Resend attack
\end{enumerate}

Assuming that there is an external attacker Eve carry out the
intercept-and-resend attack, she intercepts the quantum state $\left| \Phi  \right\rangle _j$ during the transmission
of ${\rm Bob_k}$ and ${\rm Bob_{k+1}}$, and she retransmits her own forged particle, where
$1 \le k \le m - 1$.

Since the measurement basis in this scheme $\sum\limits_{i = 0}^k {q_i }  = q_0  + q_1  +  \cdots  + q_k$ is not
published to the public, the eavesdropper Eve does not know any information
about it. Therefore, Eve can only choose one of $d$ groups
related measurement basis to get the original secret. Only when the selected
basis is the real measurement basis, she can get the right measurement
results. Then she only has the possibility of $\frac{1}{d}$
successfully obtaining the number $\sum\limits_{i = 0}^k {p_i }  = S + p_1  +  \cdots  + p_k$. It can be seen that
whether Eve can succeed largely depends on $d$, and $d$ is a large odd prime number, which is the same as the success
rate of directly guessing the secret. Then Eve needs to prepare the same
quantum state to send to ${\rm Bob_{k + 1}}$. Otherwise, Alice will check
and find out it, then she will terminate the scheme. Thus intercept-and-resend attack is invalid for this scheme.

\begin{enumerate}
\item[ii)] Entanglement measurement attack
\end{enumerate}

The second attack which can be carried out by
eavesdropper Eve is entanglement measurement attack. If Eve makes an auxiliary quantum state at first, then she
performs unitary transformation $U_E$ to entangle the
auxiliary quantum state to the transmitted particles, and finally she steals
information by measuring the auxiliary particles.

Through unitary transformation $U_E$, it can be expressed as
follows:

\begin{equation}
\label{eq2}
U_E \left| k \right\rangle \left| E \right\rangle  = \sum\limits_{m = 0}^{d - 1} {a_{km} } \left| m \right\rangle \left| {\varepsilon _{km} } \right\rangle.
\end{equation}

\begin{align}
&U_E \left| {\varphi _l^{\left( j \right)} } \right\rangle \left| E \right\rangle\nonumber\\
= &U_E \left( {\frac{1}{{\sqrt d }}\sum\limits_{k = 0}^{d - 1} {\omega ^{k\left( {l + jk} \right)} } \left| k \right\rangle } \right)\left| E \right\rangle\nonumber\\
= & \frac{1}{{\sqrt d }}\sum\limits_{k = 0}^{d - 1} {\omega ^{k\left( {l + jk} \right)} } \left( {\sum\limits_{m = 0}^{d - 1} {a_{km} \left| m \right\rangle \left| {\varepsilon _{km} } \right\rangle } } \right)\nonumber\\
 = &\frac{1}{{\sqrt d }}\sum\limits_{k = 0}^{d - 1} {\sum\limits_{m = 0}^{d - 1} {\omega ^{k\left( {l + jk} \right)} a_{km} } }\nonumber\\
  &\left( {\frac{1}{{\sqrt d }}\sum\limits_{g = 0}^{d - 1} {\omega ^{ - m\left( {g + jm} \right)} \left| {\varphi _l^{\left( j \right)} } \right\rangle } } \right)\left| {\varepsilon _{km} } \right\rangle\nonumber\\
 = &\frac{1}{d}\sum\limits_{k = 0}^{d - 1} {\sum\limits_{m = 0}^{d - 1} {\sum\limits_{g = 0}^{d - 1} {\omega ^{\left( {kl + jk^2 } \right) - \left( {mg + jm^2 } \right)}\left| {\varphi _g^{\left( j \right)} } \right\rangle } } } \left| {\varepsilon _{km} } \right\rangle.
\label{eq3}
\end{align}

where $\omega  = e^{{{2\pi i} \mathord{\left/
 {\vphantom {{2\pi i} d}} \right.
 \kern-\nulldelimiterspace} d}}$, $\left| E \right\rangle$ represents the initial state
of the auxiliary system of Eve, and $\left| {\varepsilon _{km} } \right\rangle \left( {k,m = 0,1, \cdots ,d - 1} \right)$ represents the only
pure state after the auxiliary transformation $U_E$.
Therefore, the coefficients satisfy:
\[
\sum\limits_{m = 0}^{d - 1} {\left| {a_{km} } \right|^2 }  = 1, \ \
k = 0,1, \cdots ,d - 1.
\]

To prevent the error rate from increasing, Eve sets $a_{km}  = 0$,
where $k \ne m$, $k,m \in \left\{ {0,1, \cdots ,d - 1} \right\}$, thus the equations (\ref{eq2}) and
(\ref{eq3}) can be simplified as:
\begin{center}
\[U_E \left| k \right\rangle \left| E \right\rangle  = a_{kk} \left| k \right\rangle \left| {\varepsilon _{kk} } \right\rangle.\]

\[U_E \left| {\varphi _l^{\left( j \right)} } \right\rangle \left| E \right\rangle  = \frac{1}{d}\sum\limits_{k = 0}^{d - 1} {\sum\limits_{g = 0}^{d - 1} {\omega ^{k\left( {l - g} \right)} a_{kk} \left| {\varphi _g^{\left( j \right)} } \right\rangle \left| {\varepsilon _{kk} } \right\rangle } }.\]
\end{center}

Similarly, Eve can get the following equations:
\[
\sum\limits_{k = 0}^{d - 1} {\omega ^{k\left( {l - g} \right)} a_{kk} \left| {\varepsilon _{kk} } \right\rangle }  = 0.
\]

where $g \in \left\{ {0,1, \cdots ,d - 1} \right\}$, $g \ne l$. It can get
$d$ equations for arbitrary $l \in \left\{ {0,1, \cdots ,d - 1} \right\}$.
$a_{00} \left| {\varepsilon _{00} } \right\rangle  = a_{11} \left| {\varepsilon _{11} } \right\rangle  =  \cdots  = a_{d - 1,d - 1} \left| {\varepsilon _{d - 1,d - 1} } \right\rangle$ can be calculated by $d$ equations.

In order to steal effective information, we assume that Eve performs unitary
operation $U_E$, that is

\begin{equation}
\begin{aligned}
 U_E \left| {\varphi _0^{\left( 0 \right)} } \right\rangle =&\frac{1}{{\sqrt d }}\left({a_{00} \left| 0 \right\rangle \left| {\varepsilon _{00} } \right\rangle}\right) \\
 &+\frac{1}{{\sqrt d }}\left({a_{11} \left| 1 \right\rangle \left| {\varepsilon _{11} } \right\rangle}\right)+ \cdots \\
&+\frac{1}{{\sqrt d }}\left({a_{d-1,d-1} \left| d-1 \right\rangle \left| {\varepsilon _{d-1,d-1} } \right\rangle}\right) \\
= &\frac{1}{{\sqrt d }}\left( {\left| 0 \right\rangle  + \left| 1 \right\rangle  +  \cdots  + \left| {d - 1} \right\rangle } \right)\\
&\otimes \left( {a_{00} \left| {\varepsilon _{00} } \right\rangle } \right)\nonumber
\end{aligned}
\end{equation}

Therefore, no matter what kind of quantum state is adopted, Eve can only get
the same information from the auxiliary particles. So the entanglement
measurement attack cannot be successful in this scheme.

\begin{enumerate}
\item[1)] Internal attack
\end{enumerate}

Because the conspiracy of participants is a kind of destructive attack which
is easier to steal effective information than the external attack, the
participants honest or not is related closely to the security of the scheme.

\begin{enumerate}
\item[i)] Forgery attack
\item[a)] We can assume that the participants ${\rm Bob_1}$ is an internal fraudster. After he receives the quantum state $\left| \Phi  \right\rangle _0$ from Alice, because $j = q_0$ is arbitrary number of the measurement basis $\left\{ {\left| {\varphi _l^{\left( j \right)} } \right\rangle } \right\}_l$, the probability of choosing the right measurement basis is $\frac{1}{d}$, where $d$ is a large odd prime number, and the probability of failure is $\frac{{d - 1}}{d}$, which is the same as the probability of directly guessing the secret $S$. In case of ${\rm Bob_1}$ forges quantum state transmission, Alice will find the existence of the internal fraud through random check ${\rm Bob_{u}}$. If the former participants $u\left( {1 \le u \le m} \right)$ can be checked in turn through the unitary inverse operation to find the internal fraudsters, the scheme will be terminated and remove ${\rm Bob_1}$ from the participants.
\item[b)] We can assume that the participants ${\rm Bob_j} $

\noindent $\left( {2 \le j \le m - 1} \right)$ is an internal fraudster. $j = \sum\limits_{i = 0}^j {q_i }$ is arbitrary number of the measurement basis $\left\{ {\left| {\varphi _l^{\left( j \right)} } \right\rangle } \right\}_l$. The probability of choosing the right measurement basis is $\frac{1}{d}$, where $d$ is a large odd prime number, and the probability of failure is $\frac{{d - 1}}{d}$. And he can steal the information $\sum\limits_{i = 0}^j {p_j }  = S + p_1  +  \cdots  + p_j$, thus he cannot reconstruct the secret. In case of
    ${\rm Bob_j}$ forges quantum state transmission, Alice will find the existence of the internal fraud through random check ${\rm Bob_u}$. If the former participants $u\left( {1 \le u \le m} \right)$ can be checked in turn through the unitary inverse operation to find the internal fraudsters, the scheme will be terminated and remove $
{\rm Bob_j}$ from the participants.
\item[c)] We can assume that the participants ${\rm Bob_m}$ is an internal fraudster. Because ${\rm Bob_m}$ simultaneously public transmits $R$ to Alice and ${\rm Bob_i}$ through the secure channel, Alice will immediately discover and remove the forgeries once she find $R$ is wrong.
\item[ii)] Conspiracy attack

\ \ \ \  In this scheme, it is assumed that the worst case scenario is that only the
distributor Alice and one participant are trusted. We can assume that
${\rm Bob_1}$ is honest, the remaining $m-1$
participants may carry out conspiracy attack. In the process:
\item[a)] If the dishonest participants forge particles, Alice will check and find out it.
\item[b)] If one or more participants send the wrong $p_j \left( {j \ne 1} \right)$ and $R$ to the other participants in the secret sharing phase, because they simultaneously public transmit, Alice will immediately discover it, terminate the scheme, and remove the forgeries.
\item[c)] Since the secret must be reconstructed through $S = p_0  = R - \sum\limits_{i = 1}^m {p_m }$, $m-1$ participants cannot reconstruct the secret.
\end{enumerate}

Therefore, the forgery attack and conspiracy attack of participants cannot obtain the secret in this scheme, and Alice will discover it and remove the forgeries.

\section{Summary} \ \ \ \

In this paper, a VTQSS scheme using $d$ dimension single bit quantum state is proposed. In the scheme design, we combine the binary asymmetric polynomials of the classical part with the unitary opration of the quantum part based on the mutually unbiased bases. Thus the security is guaranteed at every stage.
The distributor prevents the internal attack through random detection. Once it is found out, she will detect and eliminate the internal fraudsters by unitary inverse operation. And the participants fraud is avoided through the method of simultaneous public transmission. In addition, the security of the scheme is analyzed. Of course, due to the current technology, the complexity of the scheme still needs to be improved. We hope to propose a better verifiable quantum secret sharing scheme in the future.

\subsection*{Acknowledements} \ \ \ \
We would like to thank anonymous review for valuable comments. This work is supposed by the National Natural Science Foundation of China under Grant No.11671244.
%\begin{figure}
%\resizebox{0.75\textwidth}{!}{%
  %\includegraphics{leer.eps}
%\vspace{5cm}       % Give the correct figure height in cm
%\end{figure}
%\begin{figure*}
%\vspace*{5cm}       % Give the correct figure height in cm
%\end{figure*}
%\begin{table}
%\begin{tabular}{lll}
%\hline\noalign{\smallskip}
%\end{tabular}
% Or use
%\vspace*{5cm}  % with the correct table height
%\end{table}
%
% The section below may be edited at your convenience to acknowledge
% each author's contribution to the manuscript.
% You may remove it if you are a single author.
%

\subsection*{Authors contributions} \ \ \ \
Li-Juan Liu and Zhi-Hui Li proposed the initial idea
for this paper. Li-Juan Liu wrote the initial draft of the
manuscript, and all authors participated in the discussion
and revision of the manuscript.
% BibTeX users please use
% \bibliographystyle{}
% \bibliography{}

\begin{thebibliography}{}
\bibitem{RefA}
A. Shamir, Commun. ACM \textbf{22}, 612 (1979)
\bibitem{RefB}
 G.R. Blakley, \emph{Proceedings of the National Computer Conference} (1979), pp. 313-317
\bibitem{RefC}
 W.A. Jackson, K.M. Martin, C.M. O¡¯keefe, J. Crypt. \textbf{9}, 233 (1996)
\bibitem{RefD}
C.F. Hsu, Q. Cheng, X. Tang et al., Inf. Science. \textbf{181}, 1403 (2011)
\bibitem{RefE}
R. Bitar, E.R. Salim, IEEE Trans. Inf. Theor. \textbf{64}, 933 (2018)
\bibitem{RefF}
B. Chor, S. Goldwasser, S. Micali, B. Awerbuch, \emph{Proceedings of 26th IEEE Symposium on Foundations of Computer Science} (1985), pp. 383-395.
\bibitem{RefG}
 P. Feldman, \emph{Proceedings of 28th IEEE Symposium on Foundations of Computer Science} (1987), pp. 427-437
\bibitem{RefH}
M. Stadler, EUROCRYPT¡¯96 \textbf{1070}, 190 (1996)
\bibitem{RefI}
 P.M. Shor, \emph{Proceedings of the 35th Annual Symposium of Foundation of Computer Science} (1994)
\bibitem{RefJ}
M. Hillery, V. Buzek, A. Berthiaume. Phys. Rev. A. \textbf{59}, 1829 (1999)
\bibitem{RefK}
F.G. Deng, X.H. Li, H.Y. Zhou, Z.J. Zhang, Phys. Rev. A. \textbf{72}, 044302 (2005).
%\bibitem{RefL}
%Y.G. Yang, Q.Y. Wen, Chinese Phys. B. \textbf{17}, 419 (2008)
\bibitem{RefL}
Y. Sun , Q.Y. Wen, F.C. Zhu, Commun. Theor. Phys. \textbf{54}, 89 (2010)
\bibitem{RefM}
M. H. Dehkordi, E. Fattahi, Quantum Inf. Process. \textbf{12}, 1299 (2013)
\bibitem{RefN}
R. Cleve, D. Gottesman, H.K. Lo, Phys. Rev. Lett. \textbf{83}, 648 (1999)
\bibitem{RefO}
C.M. Bai, Z.H. Li, C.J. Liu, Y.M. Li., Euro. Phys. J. D. \textbf{71}, 1 (2017)
\bibitem{RefP}
C.M. Bai, Z.H. Li, C.J. Liu, Y.M. Li., Euro. Phys. J. D. \textbf{72}, 1 (2018)
\bibitem{RefQ}
P. Khakbiz, M. Asoudeh, Quantum Inf. Process. \textbf{18}, 1 (2019)
\bibitem{RefR}
A. Tavakoli, I. Herbauts, M. Zukowski, M. Bourennane, Phys. Rev. A. \textbf{92}, 1(2015)
\bibitem{RefS}
 V. Karimipour, M. Asoudeh, Phys. Rev. A. \textbf{92}, 030301 (2015)
\bibitem{RefT}
C.M. Bai, Z.H. Li, Y.M. Li, Commun. Theor. Phys. \textbf{69}, 513 (2018)
\bibitem{RefU}
 P.K. Sarvepalli, A. Klappenecker, Phys. Rev. A. \textbf{80}, 022321 (2009)
\bibitem{RefV}
C.M. Bai, Z.H. Li, C.J. Liu, Y.M. Li. Quantum Inf. Process. \textbf{16}, 304 (2017)
\bibitem{RefW}
K. Senthoor, P.K. Sarvepalli, Phys. Rev. A. \textbf{100}, 052313 (2019)
\bibitem{RefX}
Y.G. Yang, Y.W. Teng, H.P. Chai, Q.Y. Wen, Quantum Inf. Process. \textbf{50}, 792 (2011)
\bibitem{RefY}
X. Song, Y. Liu, Quantum Inf. Process. \textbf{15}, 851 (2016)
\bibitem{RefZ}
H. Qin, Y. Dai, Inf. Process. Lett. \textbf{116}, 351(2016)
\bibitem{Refa}
N. Hadisukmana, R.Roestam, \emph{2019 ICSECC} (2019)
\bibitem{Refb}
 C. Lu, F. Miao, J. Hou, K. Meng, Quantum Inf. Process. \textbf{17}, 310 (2018)
\bibitem{Refc}
I.D. Ivonovic, J. Phys. A: Math. Ge. \textbf{14}, 3241 (1981)
\bibitem{Refd}
W.K. Wotters, B.D. Fields, Anna. Phys. \textbf{191}, 363 (1989)
% etc
\end{thebibliography}

\end{document}